\documentclass{article}
\title{Speculations on a Unified Theory of Matter and Mind\footnote
{Based on the talk: Is Consciousness An Accident? in the
International Conference: {\em Science and Metaphysics: A
Discussion on Consciousness and Genetics} at Bangalore, India,
(June 2001).}}
\author{Manoj K. Samal,}
\date{S. N. Bose National Centre for Basic Sciences, JD/III, Salt
Lake City,\\ Kolkata 700 098, India; {\tt mks@bose.res.in}}

\begin{document}
\maketitle
\begin{abstract}
Physics is so successful today in understanding the nature of
matter. What can it say about mind ? Is it possible to have a
unified theory of matter and mind within the framework of modern
science ? Is Consciousness an accident or is it a natural
consequence of laws of nature ? Are these laws of nature the same
as the laws of physics ? We make an attempt here to unify mind
with matter based on an extended formalism borrowed from Quantum
theory where information plays a more fundamental role than matter
or thought.

\end{abstract}
\section{Introduction}

Most of the physicists today believe that we are very close to a
unified theory of matter (UTM) based on p-brane theory (a variant
of String [1-brane] theory) [1]. This theory has been well
motivated by the successes (from microchips to satellites) and the
difficulties (the presence of infinities) of Quantum Gauge Field
Theories (QED, Electro-Weak theory, GUT and SUSY GUT etc.) most of
which are based on solid experimental evidences (measuring Lande
g-factor of electron up to 10 significant places, prediction of Z
boson etc.). This `theory of everything ` tells us that all carbon
atoms (after they are born out of nucleosynthesis in the core of
stars) in this universe are the same in space and in time, and
thus is unable to explain why the carbon atoms present in a lump
of roughly three pound of ordinary matter the human brain -- give
rise to such ineffable qualities as feeling, thought, purpose,
awareness and free will that are taken to be evidence for
`consciousness'!

Is Consciousness an accident caused by random evolutionary
processes in the sense that it would \textit{not }evolve again had
the present universe to undergo a `big crunch' to start with
another `big bang'? NO! It is a fundamental property that emerges
as a natural consequence of laws of nature. Are these laws of
nature different from laws of physics? Is there a way to expand
the UTM to incorporate consciousness? In this paper I make an
attempt to point out a possible direction for such an expansion to
achieve a unified theory of matter and mind by considering
\textit{`information' to be more fundamental than matter and
energy}.

Nature manifests itself not only at the gross level of phenomena
(accessible to direct senses) but also at the subtle level of
natural laws (accessible to `refined ` senses).
\textit{Consciousness is the ability to access nature at both
these levels}. Hence everything in nature is conscious, but there
is a hierarchy in the level of consciousness. Although animals,
plants and few machines today can access to the gross level of
nature, access to the subtle level seems to be purely human. In
this sense a layman is less conscious compared to a scientist or
an artist. (Is it possible that a level of consciousness exists
compared to which a scientist or an artist of today may appear a
layman?) \textit{Once everything (both matter and mind) is reduced
to information it is possible to define} \textit{consciousness as
the capability to process information}. Because any form of our
access to nature is based on processing information with varying
degree of \textit{complexity}. The words process and complexity
will be defined in the following. Now I proceed to elaborate the
claims (italicized above) but due to the constraint of printing
space I will be brief here and the details of this scheme will be
given elsewhere.

\section{Matter}
\subsection{Phenomena}

Some animals like dogs (in listening to ultrasound) and bats (in
sensing prey through echo technique) are better equipped than
humans when it comes to direct sense experiences. But unlike
animals, humans have found out methods to extend their sense
experiences through \textit{amplification devices} like
telescopes, microscopes etc. The summary [2] of such sense
experiences (direct and extended) acquired over the last five
hundred years (equivalent to one second in a time-scale where the
age of the universe is equal to a year) is that our universe
extends in space from the size of an electron (10$^{\rm{}17}$ cm
as probed at LEP colliders at CERN, Geneva) to size of galactic
super-clusters (10$^{\rm{}30}$ cm as probed by high redshift
measurements). The theoretical possibility of Planck length
(10$^{\rm{}33}$ cm) allows for further extension in space.

Our grand universe extending over more than 60 orders of magnitude
in space has been constantly \textit{changing} over the last 12
billion years! Living systems seem to have evolved out of
non-living systems. Consciousness seems to have evolved out of
living systems.  Is the dramatic difference between animate and
inanimate, conscious and unconscious simply a difference in their
ability to process information? At the level of phenomena nature
is so vast and diverse compared to the size and comprehension of
human beings that one wonders how the collective inquiring human
minds over the centuries could at all fathom the unity behind this
diversity. This simply testifies the triumph of human mind over
the physical limitations of its body. Because the human mind is
capable of reaching a synthesis (an advanced form of information
processing) based on careful observations of diverse phenomena.

\subsection{Inanimate}
\subsubsection{Manifold}

Till Einstein, everybody thought `absolute space' is the arena
(mathematically, a manifold) on which things change with `absolute
time'. In special theory of relativity (STR) he redefined the
manifold to be flat space-time (3 + 1) by making both space and
time relative with respect to inertial observers but keeping
space-time (SpT) absolute. In general theory of relativity (GTR)
he propounded this manifold to be curved space-time that can act
on matter unlike the flat space-time. Then the Quantum Theory (QT,
the standard version, not the Bohmian one) pointed out this
manifold to be an abstract mathematical space called Hilbert space
since the space-time description of quantum processes is not
available. Quantum Field Theory (QFT) that originated in a
successful merge of QT and STR requires this manifold to be the
Quantum Vacuum (QV). Unlike the ordinary vacuum, QV contains
infinite number of `virtual' particles that give rise to all
matter and interactions [3]. Every quantum field has its own QV
and a successful amalgamation of QT and GTR (called Quantum
Gravity, and yet to be achieved) may connect the curved space-time
with the QV of gravitational field. Finally, String (or p-brane)
theory demands all fundamental entities to be strings (or
p-branes) in ten-dimensional space-time. This evolution in our
understanding clearly shows the necessity and importance of a
background manifold to formulate any scientific theory.

\subsubsection{Basic Constituents}

Energy and matter were considered to be the two basic constituents
of our universe till Einstein showed their equivalence through
E=mc$^{\rm{}2}$. All forms of energy are interconvertible. Matter
in all its forms (solid, liquid, gas and plasma) consists of
atoms. The simplest of all atoms is the hydrogen atom that
contains an electron and a proton. If one considers the (now
outdated) Bohrian picture of hydrogen atom being a miniature solar
system where the electron revolves around the proton in circular
orbit then one realizes that most of the hydrogen atom is empty
space. This means that if one blows up the hydrogen atom in
imagination such that both electron and proton acquire the size of
a football each then they need to be separated by 100 km or more!
Hence one would think that even if 99.99 \% of the hydrogen atom
consists of vacuum the point-like electron and proton are
\textit{material} particles. But that is not so.

The elementary particle physics tells us that all the \textit{visible
matter} (composition of \textit{dark matter }is not yet surely known) in
the universe is made of six leptons and six quarks along with their
antiparticles. Although they are loosely called \textit{particles} they
are not like the ordinary particles we experience in daily life. A
`classical' particle is \textit{localized} and \textit{impenetrable}
whereas a `classical' wave can be extended from minus infinity to plus
infinity in space and many wave modes can simultaneously occupy the same
space (like inside the telephone cable). But with the advent of quantum
mechanics this seemingly contradictory differences between particle and
wave lost their sharpness in the quantum world. A quantum object can
simultaneously `be' a particle and a wave until a measurement is made on
it. According to QFT all fundamental entities are quantum fields (not
material in the conventional sense) that are neither particle nor wave in
the classical sense.

\subsubsection{Evolution}

Despite various specializations, all physical sciences share a
common goal: given the complete specification of a physical system
at an initial time (called the initial conditions) how to predict
what will it be at a later time. To predict with exactness one
needs the accurate initial conditions and the laws that govern the
evolution of the system (called the dynamical laws). Either the
lack of exact specifications of initial conditions or the
intractability of huge number of equations (that express dynamical
laws) can lead to a probabilistic description rather than a
deterministic one. However, there are \textit{chaotic} systems
that can be both deterministic yet unpredictable because of their
extreme sensitivity to initial conditions. Apart from dynamical
laws there are other laws like E=mc$^2$ etc., which do not involve
time explicitly. Could there be laws at the level of initial
conditions that guide us to choose a particular set over another?

A physical law is like the hidden thread (unity) of a garland with various
flowers representing diverse natural phenomena. Its character seems to
depend on characteristic scales (denoted by \textit{fundamental constants}
like Planck length, Planck's constant, and speed of light etc). Why should
there be different set of laws at different scales? Most physicists
believe that quantum mechanics is universal in the applicability of its
laws like Schr{\footnotesize{}Ö}dinger's equation and classicality of the
everyday world is a \textit{limiting case}. But there exists no consensus
at present regarding the \textit{emergence} of this limiting case.

\subsubsection{Guiding Principles}

How does one formulate these physical laws? The principle of
\textit{relativistic} \textit{causality} helps. Do physical laws
change? Is there a unity behind the diversity of laws? Is it
possible to understand nature without laws? The concept of
symmetry (invariance) with its rigorous mathematical formulation
and generalization has guided us to know the most fundamental of
physical laws. Symmetry as a concept has helped mankind not only
to define `beauty' but also to express the `truth'. Physical laws
tries to quantify the truth that appears to be `transient' at the
level of phenomena but symmetry promotes that truth to the level
of `eternity'.

\subsubsection{Interactions}

The myriad mosaic of natural phenomena is possible because, not only each
fundamental entity evolves with time but also it can interact with the
other basic constituents. All the physical interactions (known so far) can
be put into four categories: (i) gravitational interaction (the force that
holds the universe), (ii) electromagnetic interaction (the force that
holds the atom, and hence all of us), (iii) strong nuclear interaction
(the force that holds the nucleus), and (iv) weak nuclear interaction (the
force that causes radioactive decay). At present (ii) and (iv) are known
(observationally) to be unified to a single force called ElectroWeak.
Grand unified theories (GUT) and their extensions (SUSY GUT) for (ii),
(iii) and (iv) do exist. Ongoing research aims to unify (i) with such
theories.

According to QFT the basic matter fields interact by exchanging messenger
fields (technically called \textit{gauge fields}) that define the most
fundamental level of \textit{communication} in nature. Both matter fields
and gauge fields originate in the fluctuations of QV and in this sense
everything in universe including consciousness is, in principle, reducible
to QV and its fluctuations [4]. Communication in nature can happen either
via the \textit{local} channel mediated by gauge fields or by the
\textit{nonlocal} EPR [5] type channels (through entanglement) as was
demonstrated by recent \textit{quantum teleportation} experiments.

\subsubsection{Composite Systems}

Till the importance of {\it quantum entanglement} was realized in
recent times the \textit{whole} was believed to be \textit{just}
the sum of \textit{parts}. But the whole seems to be much more
than just the sum of parts in the `quantum' world as well as
classical systems having \textit{complexity}. It makes
\textit{quantum entanglement} a very powerful resource that has
been utilized in recent times for practical schemes like quantum
teleportation, quantum cryptography and quantum computation.
Quantum nonlocality indicates that the universe may very well be
\textit{holographic} in the sense that \textit{the whole} is
reflected in \textit{each part}.

\subsection{Animate}
\subsubsection{Manifold}

Space-time (3 + 1) is the manifold for all biological functions at
the phenomenal level that can be explained by classical physics.
If one aims to have a quantum physical explanation of certain
biological functions then the manifold has to be the Hilbert
Space.

\subsubsection{Basic Constituent}

Cell is the basic constituent of life although the relevant information
seems to be coded at the subcell (genetic) level. Neuron (or Microtubules
and Cytoskeletons) could be the physical substratum of brain depending on
classical (or quantum) viewpoint.

\subsubsection{Evolution}

Does biological evolution happen with respect to the physical time? If
yes, then will the physical laws suffice to study biological evolution in
the sense they do in chemistry? If no, then is the biological
\textit{arrow of time} different from the various arrows of physical time
(say, cosmological or the thermodynamical arrow of time)? Is there a need
for biological laws apart from physical laws to understand the functioning
of biological systems?

\subsubsection{Guiding Principles}

\textit{Survivability} is the guiding principles in biological systems.
Organisms constantly adapt to each other through evolution, and thus
organizing themselves into a delicately tuned ecosystem.  Intentionality
may also play a very important role in the case of more complex
biosystems.

\subsubsection{Interactions}

The interaction occurs by exchange of chemicals, electric signals,
gestures, and language etc. at various levels depending upon the level of
\textit{complexity} involved.

\subsubsection{Composite Systems}

Composite systems are built out of the basic constituents retaining the
relevant information in a holographic manner. The genetic information in
the zygote is believed to contain all the details of the biology to come
up later when the person grows up. The genes in a developing embryo
organize themselves in one way to make a liver cell and in another way to
make a muscle cell.

\subsection{Discussions}
\subsubsection{How `material' is physical?}

Anything that is \textit{physical} need not be `material' in the sense we
experience material things in everyday life.  The concept of energy is
physical but not material. Because nobody can experience energy directly,
one can only experience the manifestations of energy through matter.
Similarly the concept of a `classical field' in physics is very abstract
and can only be understood in terms of analogies. Still more abstract is
the concept of a `quantum field' because it cannot be understood in terms
of any classical analogies. But at the same time it is a wellknown fact in
modern physics that all \textit{fundamental entities} in the universe are
\textit{quantum fields}. Hence one has to abandon the prejudice that
anything `physical' has to be `material'.

\subsubsection{Is reductionism enough?}

The reductionist approach:  observing a system with an increased
resolution in search of its basic constituents has helped modern
science to be tremendously successful. The success of modern
science is the success of the experimental method that has reached
an extreme accuracy and reproducibility. But the inadequacy of
reductionism in physical sciences becomes apparent in two cases:
{\it emergent phenomena} and {\it quantum nonlocality}. Quantum
nonlocality implies a holographic universe that necessitates a
\textit{holistic} approach [6].

Though it is gratifying to discover that everything can be traced
back to a small number of quantum fields and dynamical laws it
does not mean that we now understand the origin of earthquakes,
weather variations, the growing of trees, the fluctuations of
stock market, the population growth and the evolution of life?
Because each of these processes refers to a system that is
\textit{complex}, in the sense that a great many independent
agents are interacting with each other in a great many ways. These
complex systems are \textit{adaptive} and undergo
\textit{spontaneous self-organization }(essentially nonlinear)
that makes them \textit{dynamic} in a qualitatively different
sense from \textit{static} objects such as computer chips or
snowflakes, which are merely complicated. Complexity deals with
\textit{emergent phenomena.} The concept of complexity is closely
related to that of \textit{understanding, }in so far as the latter
is based upon the accuracy of\textit{ model }descriptions of the
system obtained using condensed information about it [7].

In this sense there are three ultimate frontiers of modern
physics: the very small, the very large and the very complex.
Complex systems cease to be merely complicated when they display
coherent behaviour involving \textit{collective organization} of
vast number of degrees of freedom. Wetness of water is a
collective phenomenon because individual water molecules cannot be
said to possess wetness. \textit{Lasers, super-fluidity and
super-conductivity} are few of the spectacular examples of
\textit{complexity} in macroscopic systems, which cannot be
understood alone in terms of the microscopic constituents. In
every case, groups of entities seeking mutual accommodation and
self-organization somehow manage to transcend the individuality in
the sense that they acquire collective properties that they might
never have possessed individually. In contrast to the linear,
reductionist thinking, complexity involves \textit{nonlinearity}
and \textit{chaos} and we are at present far from understanding
the complexity in inanimate processes let alone the complexity in
living systems.

\subsubsection{Emergence of Life}

Is life nothing more than a particularly complicated kind of carbon
chemistry? Or is it something subtler than putting together the chemical
components? Do computer viruses have life in some fundamental sense or are
they just pesky imitations of life? How does life emerge from the
quadrillions of chemically reacting proteins, lipids, and nucleic acids
that make up a living cell? Is it similar to the emergence of thought out
of the billions of interconnected neurons that make up the brain? One hope
to find the answer to these questions once the dynamics of
\textit{complexity} in inanimate systems is well understood.

\section{Mind}
\subsection{Phenomena}

Mind is having three states: awake, dream, dream-less sleep. Mind
is capable of free-will, self-perception (reflective) and
universal perception (perceptual) in its `awake' state. Can it be
trained to have all these three attributes in the states of dream
and dream-less sleep? Can there be a fourth state of mind that
transcends all the above three states? Where do the brain end and
the mind begin? Due to its \textit{global} nature, mind cannot lie
in any particular portion of the brain. Does it lie everywhere in
the brain? This would require nonlocal interactions among various
components of the brain. If there were no such nonlocal
communication then how does the mind emerge from the brain?

Can anybody think of anything that transcends space-time? Is mind
capable of thinking something absolutely new that has not been
experienced (directly or indirectly) by the body? Nobody can think
of anything \textit{absolutely} new. One can only think of a new
way of arranging and/or connecting things that one has ever
learnt. In this sense intellect is constrained by reason whereas
imagination is not. But imagination is not acceptable to intellect
unless it is logically consistent with what is already known.
Imagination helps to see a new connection but intellect makes sure
that the new connection is consistent with the old structure of
knowledge. This is the way a new structure in knowledge is born
and this process of acquiring larger and larger structure (hence
meaning or synthesis) is the \textit{learning} process.  Science
is considered so reliable because it has a stringent methodology
to check this consistency of imagination with old knowledge.

Can one aspire to study the mind using methodology of (physical) sciences?
Seeing the tremendous success of physical sciences in the external world
one would think its methodology to work for understanding the inner world.
It is not obvious \textit{a priori} why should not QT work in this third
ontology when it has worked so successfully with two different ontologies?
We aim to understand nature at a level that transcends the inner and the
outer worlds by synthesizing them into a more fundamental world of quantum
information.
\subsection{Formalism}
\subsubsection{Manifold}

A physical space-time description of mind is not possible because
thoughts that constitute the mind are \textit{acausal}: it does
not take 8 minutes for me to think of the sun although when I look
at the sun I see how it was 8 minutes ago. We will assume that it
is possible to define an abstract manifold for the space of
thoughts (say, T-space). An element of T-space is a thought-state
(T-state) and the manifold allows for a continuous change from one
T-state to another. I presume that T-space is identical with mind
but it need not be so if mind can exist in a thoughtless but awake
state, often called {\it turiya}.
\subsubsection{Basic Constituent}

How does one define a T-state?  That requires one to understand
what is a `thought'? A thought always begins as an idea (that
could be based on self and universal perception) and then
undergoes successive changes in that idea but roughly remaining
focused on a theme. Change from one theme to another is triggered
by a new idea. Hence I would suggest that the basic constituent of
T-state is \textit{idea.} An idea is like a `snapshot' of
experience complete with all sense data whereas a thought
(T-state) is like an \textit{ensemble} (where each element is not
an exact replica of the other but has to be very close copy to
retain the focus on the theme) of such snapshots.
\subsubsection{Evolution}

There are two types of evolution in T-space. First, the way an
ensemble of ideas evolves retaining a common theme to produce a
thought. To \textit{concentrate} means to linger the focus on that
theme. This evolution seems to be nonlinear and nondeterministic.
Hence the linear unitary evolution of QT may not suffice to
quantify this and will be perhaps best described in terms of the
mathematics of \textit{self-organization} and
\textit{far-from-equilibrium phenomena}. [On the practical side
the time-tested techniques of Yoga teach us how to linger the
focus on a theme through the practice of \textit{dharana, dhyan
and samadhi.} ]\textit{ }The second type involves a change from
one particular thought into another and this evolution could be
linear and perhaps can be calculated through a \textit{probability
amplitude} description in the line of QT. Given the complete
description of a thought at an initial time the refutability of
any theory of mind amounts to checking how correctly it can
predict the evolution of that thought at a later time.
\subsubsection{Guiding Principles}

If the guiding principle for evolution in biological world is
\textit{survivability} then in T-space it is
\textit{happiness-ability}. A constant pursuit of happiness
(although its definition may vary from person to person) guides
the change in a person's thoughts. Each and every activity (begins
as mental but may or may not materialize) is directed to procure
more and more happiness in terms of sensual pleasures of the body,
emotional joys of the imagination and rational delights of the
intellect.
\subsubsection{Interactions}

Can a thought (mind) interact with another thought (mind)? Can
this interaction be similar to that between quantum fields?
Perhaps yes, only if both thought and quantum fields can be
reduced to the same basic entity. Then it will be possible for
thought (mind) to interact with matter.  What will be the
messenger that has to be exchanged between interacting minds or
between interacting mind and matter? This ultimate level of
communication has to be at the level of QV and hence it may amount
to \textit{silence} in terms of conventional languages.  But can
any receiver (either human mind or any other mind or equipment) be
made so sensitive to work with this ultimate level of
communication? Interaction with the environment is believed to
\textit{decohere} a quantum system that causes the emergence of
classicality in physical world. A completely isolated system
remains quantum mechanical. Can a completely isolated mind exhibit
quantum mechanical behaviour in the sense of
\textit{superposition} and \textit{entanglement?}
\subsubsection{Composite Systems}

In the T-Space a thought is an ensemble of ideas and a mind-state
is composed of thoughts. Behaviour, feeling and knowledge of self
and universe are in principle reducible to composite subsets in
the T-space.
\subsection{Discussions}
\subsubsection{Working definition of Consciousness}

Consciousness (at the first level) is related to one's
\textit{response} $R$ to one's environment. This response consists
of two parts: \textit{habit} $H$ and \textit{learning} $L$. Once
something is learnt and becomes a habit it seems to drop out of
consciousness. Once driving a bicycle is learnt one can think of
something else while riding the bicycle. But if the habit changes
with time then it requires conscious attention. We have defined
\textit{learning} earlier as a process to find commensurability of
a new experience with old knowledge. One has to learn anew each
time there is a change in the environment. Hence consciousness is
not the response to the environment but is the \textit{time of
rate of change of the response} i.e.
\begin{equation}
C = dR/dt \end{equation} where, $R = H + L$.

The hierarchy in consciousness depends on the magnitude of this
time derivative. Everything in the universe can be fit in a scale
of consciousness with unconscious and super-conscious as the limit
points. It is obvious that all animals show response to their
environment, so does some of the refrigerators, but there is a
hierarchy in their response. Through the use of `cresco-graph' and
`resonant cardio-graph' of J. C. Bose one can see the response of
botanical as well as inanimate world. We cannot conclude that a
stone is unconscious just because we cannot communicate with it
using our known means of communication. As technology progresses,
we will be able to \textit{measure} both the response function
\textit{R} and its time derivative. If this is the definition what
can it tell about the future evolution of humans? My guess is that
we would evolve from conscious to super-conscious in the sense
that \textit{genes will evolve to store the cumulative learning of
the human race.}
\subsubsection{Emergence of Consciousness}

The first step in understanding consciousness consists of using
reductionist method to various attributes of consciousness. A
\textit{major} part of the studies done by psychologists (and
their equivalents doing studies on animals) and neuro-biologists
falls under this category.  Such studies can provide knowledge
about mind states (say M1, M2,\ldots) but cannot explain the
connection between these mind states with the corresponding brain
states (say B1, B2, \ldots). Because this kind of dualistic model
of Descarte would require to answer a) where is mind located in
the brain, and b) if my mind wants me to raise my finger, how does
it manage to trigger the appropriate nerves and so on in order for
that to happen without exerting any \textit{known forces} of
nature?

To find out how the mind actually works one needs to have a
\textit{theory of mind}, that will relate the sequence of mental
states M1, M2, M3,\ldots by providing laws of change (the
dynamical laws for the two types of evolution discussed above)
that encompass the mental realm after the fashion of the
\textit{theory of matter} that applies to the physical realm, with
its specific laws. Such a theory of mind is possible if we
synthesize the results of studies on attributes of consciousness
to define the exact nature of the manifold and the basic entities
of the T-space (or Mind-Space). Once this is achieved then one can
attempt to explain the \textit{emergence of consciousness} taking
clues from \textit{complexity theory} in physical sciences. But
such an extrapolation will make sense provided both M-states and
B-states can be reduced to something fundamental that obeys laws
of complexity theory. We propose in the next section that
information is the right candidate for such a reduction.
\subsubsection{Role of Indian Philosophy (IP)}

(1) Unlike the Cartesian dichotomy of mind and body some schools
of IP like \textit{Vaisheshika} and \textit{Yoga} treat both mind
and body in a unified manner. Since (western) science is based on
Cartesian paradigm it cannot synthesize mind and body unless it
takes the clue from oriental philosophies and then blend it with
its own rigorous methodology.

\noindent (2) In terms of sense awareness, awake, dream and
dream-less sleep states are often called as conscious,
subconscious and unconscious states. A great conceptual step taken
by IP in this regard is to introduce a \textit{fourth}
\textit{state} of mind called \textit{turiya} that is defined to
be none of the above but a combination of all of the above states.
This state is claimed to be the super-conscious state where one
transcends the limitations of perceptions constrained by
space-time (3 +1). \textit{Patanjali} has provided very scientific
and step by step instructions to reach this fourth state through
\textit{samyama} (\textit{dharana},\textit{ dhyana} and
\textit{samadhi }are different levels of \textit{samyama).} The
scientific validity of this prescription can be easily checked by
controlled experiments. Nobody can understand the modern physics
without going through the prerequisite mathematical training. It
will be foolish for any intelligent lay person to doubt the truth
of modern physics without first undergoing the necessary training.
Similarly one should draw conclusion about yogic methods only
after disciplined practice of the eight steps of yoga.

\noindent (3) IP can provide insights regarding the role of mind
in getting happiness and thus a better understanding of mind
itself. Happiness lies in what the mind perceives as pleasurable
and hence the true essence of happiness lies in mind and not in
any external things. Once the body has experienced something mind
is capable of recreating that experience in the absence of the
actual conditions that gave rise to the experience in the first
place. One can use this capacity of mind to create \textit{misery}
or \textit{ecstasy} depending on one's ability to guide one's
mind.

\noindent (4) There is a concept of \textit{anahata nada} (the
primordial sound) in IP. Sometimes the possibility of having a
universal language (like \textit{Sandhya bhasa} etc.) to
communicate with everything in the universe is also mentioned.
Modern physics tells us that the only universal language is at the
level of gauge bosons and QV.  Is there any connection between
these two? Can a (human) mind be trained to transmit and receive
at the level of QV?

\noindent (5) It is said that whole body is in the mind whereas
the whole mind is not in the body. How does mind affect the body?
If one believes in the answer given by IP then the results
obtained in this regard by the western psychology appears to be
the tip of the iceberg only. Can science verify these oriental
claims through stringently controlled experiments?
\section{Unification}
\subsection{Information}

Information seems to be abstract and \textit{not real} in the
sense that, it lies inside our heads. But information can, not
only \textit{exist} outside the human brain (i.e. library, a CD,
internet etc.) but also can be \textit{processed} outside human
brain (i.e. other animals, computers etc.).  Imagine a book
written in a dead language, which nobody today can decipher. Does
it contain information? Yes. Information exists. It does not need
to be \textit{perceived} or \textit{understood} to exist. It
requires no intelligence to interpret it. In this sense
information is as real as matter and energy when it comes to the
internal structure of the universe [8]. But what we assume here is
that \textit{information is more fundamental than matter and
energy} because everything in the universe can be ultimately
reduced to information.

\textit{Information is neither material nor non-material}. Both,
quantum fields and thoughts can be reduced to information. If the
human mind is not capable (by the methods known at present) of
understanding this ultimate information then it is the limitation
of the human mind. This may not remain so as time progresses. The
whole of physical world can be reduced to information [9]. Is
information classical or quantum? There are enough indications
from modern physics that although it can be classical at the
everyday world it is quantum at the most fundamental level. The
quantum information may have the advantage of describing the
\textit{fuzziness} of our experiences.
\subsection{Formalism}
\subsubsection{Manifold}

The manifold is an information field (I-field) for classical
information (like that of Shannon or Fisher etc.) Hilbert space of
QT is the manifold to study quantum information. But if quantum
information has to be given an \textit{ontological} reality then
it may be necessary for the manifold to be an \textit{extended}
Hilbert space.
\subsubsection{Basic Constituent}

A \textit{bit }or a \textit{qubit} is the basic entity of
information depending on whether it is treated as classical or
quantum respectively. Information can be of two types:
\textit{kinetic} and \textit{structural}, but they are convertible
to each other [8].
\subsubsection{Evolution}

All organized systems contain information and addition of
information to a system manifests itself by causing the system to
become more organized or reorganized. The laws for evolution of
information are essentially laws of \textit{organization}. Are
these laws different from the physical laws? Is there an
equivalent in the world of information of fundamental principles
like principle of least action in physical world?
\subsubsection{Guiding Principles}

\textit{Optimization} seems to be the guiding principle in the
world of information. What gives rise to the structure in the
information such that we acquire an understanding or meaning out
of it? Is there a \textit{principle of least information} to be
satisfied by all feasible structures?
\subsubsection{Interactions}

The interaction at the level of information has to be the ultimate
universal language. What could be that language? The only
fundamental language known to us is that of the gauge fields that
communicate at the level of QV. Could the gauge fields serve as
quanta of information?  How far is this language from the
conventional language? Can this help us to communicate with not
only with other creatures incapable of our conventional language
but also with the inanimate world? Time is not yet ripe to answer
these questions.
\subsubsection{Composite Systems}

How can every composite system of information (like a gene or a
galaxy) be expressed in terms of bits or qubits? Does the
holographic principle also apply to information?

\subsection{Discussions}
\subsubsection{Consciousness and Information}

{There is no doubt that sooner or later all attributes of
consciousness can be reduced to information. This is just a matter
of time and progress in technology. That will complete the
understanding of consciousness at the gross level of phenomena but
will harbinger the understanding of consciousness at the subtle
level of laws. The synthesis of the phenomenological studies of
consciousness will be possible by treating information as the most
basic ontological entity, which can unify mind and matter. The
emergence of consciousness will be understood in terms of
nonlinear, far-from-equilibrium complex processes that lead to
spontaneous self-organization and adaptation of structures in the
manifold of quantum information. }

{Consciousness will be seen as the ability to process quantum
information in an effective way. Depending on the degree of
complexity involved the processing would encompass activities
starting from the way a planet knows which is the path of least
action to the way modern supercomputers do simulations of reality
to the way a scientist makes a discovery or an artist traps beauty
on a canvass through the nuances of truth. The limit points of
unconscious and super-conscious would correspond to the limiting
cases of no information processing and infinite information
processing respectively. Subconscious will be interpreted as
partial information processing.}

Every entity in the universe has to take a decision at every moment of
time for its existence although the word existence may mean different
things to different entities.  The chance for continuation of existence is
enhanced if the best decision on the basis of available information is
taken. This is a process of optimization and the more conscious an entity
is more is its ability to optimize.
\subsubsection{Limitations of understanding}

{Is there any fundamental principle (or theorem) that puts limit
on the understanding of both mind and matter by reducing them to
information and then applying methodology of physical sciences to
understand life and consciousness as emergent phenomena?  Since
this approach heavily relies on mathematics the limitations of
deductive logic as pointed out by G\"odel in his famous
incompleteness theorem may put the first limit. The second
constraint may come from QT if it turns out (after having
rigourous information theoretic formulation of both matter and
mind) that the information related to mind is complementary to the
information found in matter.  I personally feel that this is quite
unlikely because I believe that information at the fundamental
level cannot be dualistic.}
\section{Conclusions}

Unlike the Cartesian duality between mind and body, understanding
consciousness requires first to understand matter and mind in a
unified way. This can be achieved by giving information the most
primary status in the universe. Then a generalized theory of
quantum information dynamics has to be formulated (see the table
(in the last page): The World Matrix, for a summary). The line of
attack involves three steps: 1) understanding emergent phenomena
and complexity in inanimate systems, 2) understanding life as
emergent phenomena, and 3) understanding consciousness as emergent
phenomena. The attributes of consciousness can be understood only
by a prudent application of both reductionism and holism. But the
emergence of consciousness will be understood as an emergent
phenomenon in the sense of structural organizations in the
manifold of information to yield feasible structures through which
we attribute meaning and understanding to the world.
\section{Acknowledgement}

I am thankful to the organizers of the conference, especially
Prof. B. V. Sreekantan and Dr. Sangeetha Menon, for providing the
local hospitality and intellectual support.

\section{References}

[1] Sarkar U., 2001, \textit{New Dimensions New Hopes}; arXiv:
hep-ph/0105274

\noindent [2] Davies P., 1989,\textit{The New Physics}, Cambridge
University Press.

\noindent [3] Sreekantan B. V., 2001, this conference proceeding

\noindent [4] Sreekantan B. V., 1999, \textit{Scientific
Explanations and Consciousness} , Proceedings of the national
conference on Scientific and Philosophical Studies on
Consciousness, held at NIAS, Bangalore.

\noindent [5] Einstein A, Podolsky B., Rosen N (EPR), 1935,
\textit{Quantum theory and Measurement}, edt. By Zurek, W. H. and
Wheeler, J. A.

\noindent [6] Bohm D. and Hiley B.,1993, \textit{The Undivided
Universe}, Routledge, London.

\noindent [7] Badii, R and Politi, A, 1997, \textit{Complexity:
Hierarchical Structures and Scaling in Physics}, Cambridge
University Press.

\noindent [8] Stonier, T., 1990, \textit{Information and the
internal structure of the Universe}, SpringerVerlag.

\noindent [9] Roy Frieden B., 1999, \textit{Physics from Fisher
Information: A Unification}, Cambridge University Press. \vskip
2.5cm
\centerline{\Large \bf{The World Matrix: (Summary of various
Worlds)}} \vskip 7mm {\large
\begin{tabular}{||c||c||c||c||c||}
\hline \hline
\phantom{0.1}&\phantom{0.1}&\phantom{0.1}&\phantom{0.1}&\phantom{0.1}
\\ {\bf Character}&{\bf Physical}&{\bf Biological}&{\bf Mental}&{\bf Information}
\\ \hline
\phantom{0.1}&\phantom{0.1}&\phantom{0.1}&\phantom{0.1}&\phantom{0.1}
\\{\bf Manifold}&SpT (3 + 1),&SpT (3 + 1),&M-Space&I-Field, \\
&Hilbert Space&Hilbert Space&(Abstract&Extended \\ &QV, SpT
(10)&&Mathematical&Hilbert Space\\ &&&Space)&
\\ \hline
\phantom{0.1}&\phantom{0.1}&\phantom{0.1}&\phantom{0.1}&\phantom{0.1}\\
{\bf Basic}&Wave-function&Cell, Neuron,&Idea (based&Bit \\{\bf
Constituents} &Quantum fields&Micro-tubule&on self or&(classical)
\\ &Strings&Cyto-skeleton&universal&Qubit
\\ &p-branes&&perception&(quantum)
\\ \hline
\phantom{0.1}&\phantom{0.1}&\phantom{0.1}&\phantom{0.1}&\phantom{0.1}\\
{\bf Evolution}&Phys. Laws&Phys. Laws&Laws for& Laws for\\ &(
mostly diff. &Bio. Laws& evolution of& evolution of
\\ &equations)&&thought&organization \\ \hline
\phantom{0.1}&\phantom{0.1}&\phantom{0.1}&\phantom{0.1}&\phantom{0.1}\\
{\bf Guiding}&Symmetry&Survivability&Happiness-&Optimization
\\{\bf Principles} &(Group&Intentionality&Ability&
\\ &Theoretical)&&&
\\ \hline
\phantom{0.1}&\phantom{0.1}&\phantom{0.1}&\phantom{0.1}&\phantom{0.1}\\
{\bf Interactions}&Gravity,&Chemicals,&Primordial& Local, and \\
&Electro-weak,&Electric&Sound or& Non-local
\\ &Strong&Signals,&Vibrations&(EPR)
\\ &Nuclear&Language,&&channels
\\ &Interaction&Gesture&&
\\ \hline
\phantom{0.1}&\phantom{0.1}&\phantom{0.1}&\phantom{0.1}&\phantom{0.1}\\
{\bf Composite}&Many-body&Plants,&Thought&Complex \\ {\bf
Systems}&Systems with&Animals&(Ensemble of&Systems with
\\ &or without&&Ideas with&Hierarchy in
\\ &interactions&&ordering)&Organization
\\
\phantom{0.1}&\phantom{0.1}&\phantom{0.1}&\phantom{0.1}&\phantom{0.1}\\
\hline \hline
\end{tabular}}
\end{document}